\documentclass[12pt,preprint]{aastex}

\shorttitle{ Hydrogen lines of $\eta$ Car } 
\shortauthors{ Davidson et al. }

\begin{document}

\title{ A Change in the Physical State of $\eta$ Carinae?{\altaffilmark{1}}} 

\revised{ 2004 October 4 }

\author{ K.\ Davidson{\altaffilmark{2,3}}, J.\ Martin{\altaffilmark{3}}, 
    R.\ M.\ Humphreys{\altaffilmark{3}},  K.\ Ishibashi{\altaffilmark{4}},  
    T.\ R.\ Gull{\altaffilmark{5}},  O.\ Stahl{\altaffilmark{6}},  \\
    K.\ Weis{\altaffilmark{7}},  D.\ J.\ Hillier{\altaffilmark{8}}, 
    A.\ Damineli{\altaffilmark{9}}, M.\ Corcoran{\altaffilmark{10}}, 
    and F.\ Hamann{\altaffilmark{11}}  }

\altaffiltext{1}
  { This research is part of the Hubble Space Telescope Treasury
  Project for Eta Carinae, supported by grants GO-9420 and GO-9973 
  from the Space Telescope Science Institute, which is operated by 
  the Association of Universities for Research in Astronomy, Inc., 
  under NASA contract NAS5-26555. }
\altaffiltext{2}   
  { kd@astro.umn.edu }
\altaffiltext{3}   
  { School of Physics \& Astronomy, University of Minnesota,
    116 Church St.\ SE, Minneapolis, MN 55455. }
\altaffiltext{4}   
  { Center for Space Research, MIT, 77 Massachusetts Ave.,
    NE80-6011, Cambridge, MA 02139  }
\altaffiltext{5}   
  { NASA/Goddard Space Flight Center, Code 681,
    Greenbelt, MD 20771 }
\altaffiltext{6}   
  { Landessternwarte Heidelberg, K\"{o}nigsstuhl,
    69117 Heidelberg, Germany }
\altaffiltext{7}   
  { Astronomisches Institut, Ruhr-Universit\"{a}t Bochum, 
    Universit\"{a}tsstr.\ 150, 44780 Bochum, Germany }
\altaffiltext{8}   
  { Dept.\ of Physics \& Astronomy, University of Pittsburgh,
    3941 O'Hara St., Pittsburgh, PA 15260 }
\altaffiltext{9}   
  { IAG -- University of S\~{a}o Paulo, R.\ do Mat\~{a}o 1226,
    05508-900 S\~{a}o Paulo, Brazil }
\altaffiltext{10}  
  { NASA/Goddard Space Flight Center, Code 662, Greenbelt, MD 20771 } 
\altaffiltext{11}  
  { Dept.\ of Astronomy, University of Florida, PO Box 112055,
    Gainsville, FL 32611 }

\begin{abstract}
During $\eta$ Car's spectroscopic event in mid-2003, the 
stellar wind's bright H$\alpha$ and H$\beta$ emission lines 
temporarily had a distinctive shape unlike that reported 
on any previous occasion, and particularly unlike the 
1997--98 event.  Evidently the structure of the wind changed 
between 1997 and 2003.  Combining this with other evidence, we
suspect that the star may now be passing through a rapid stage 
in its recovery from the Great Eruption seen 160 years 
ago.  In any case the data indicate that successive spectroscopic
events differ, and the hydrogen line profiles are quantitative 
clues to the abnormal structure of the wind during a spectroscopic 
event.  
\end{abstract}


\section{ INTRODUCTION }

Systematic spectral variations are clues to the nature of $\eta$ Car 
--- our most observable very massive star dangerously close 
to the Eddington limit, and the site of a prototype ``supernova impostor'' 
eruption seen about 160 years ago.  On several occasions between 1945 and 
1990 the general excitation or ionization decreased for a few months, in 
the stellar wind and also in slow-moving ejecta located 
within 1000 AU of the star (see \citet{zanella84,dh97,ad99,
kd99a}; and other refs.\ cited therein). 
Prompted by a fresh example observed in 1992, \citet{ad96} noticed that 
these events recur in a 5.5-year cycle which had also appeared in 
photometry \citep{pw94}.  The next two spectroscopic events occurred 
near 1998.0 and 2003.5, close to the predicted times.

As noted in Section 5 below, at least three rival scenarios 
may explain an event of this type.    The star may
eject a shell, possibly triggered by a hypothetical companion 
star in a 5.5-year orbit;  the excitation of our side 
of the wind may temporarily decrease when a hot secondary star 
moves to the far side of its orbit;  UV radiation from a hot 
companion may become engulfed in the primary wind near 
periastron; or some mixture of these effects 
may occur.  Unfortunately no quantitative model has been 
developed, and one can imagine other possibilities.  
A strange detail, highly relevant to the discussion in this
paper, is that the 5.5-year cycle probably did not exist before 
the 1940's -- at least not in its present form; see Section 5 
below and references cited there.

Suitable data on these events were scarce until recently.  
Ground-based spectroscopy  of the star is seriously contaminated 
by diffuse ejecta located at radii 0.2{\arcsec} -- 2{\arcsec};  
and, so far as we know, no long-term, spatially consistent 
ground-based observations of a broad wavelength range were attempted.  
The Hubble Space Telescope (HST) can attain the required high
spatial resolution, but was used for spectroscopy of $\eta$ Car on 
only a few occasions from 1991 to 2001 -- fortunately including part of 
the 1997--98 event (see below).  The fund of available data 
greatly improved during 2002--03, when we obtained numerous HST 
observations using the Space Telescope Imaging Spectrograph (STIS) 
as part of a Hubble Treasury Project to observe the 2003.5 event 
\citep{kd02b}.

Here we report a particularly striking result:  H$\alpha$ and H$\beta$, 
the brightest observable emission lines in the stellar wind, briefly 
developed unusual velocity profiles which were {\it quite different 
from their appearance during the previous event in 1997--98.\/}  
This development, combined with other circumstances noted in Sections
4 and 5 below, leads us to suspect that $\eta$ Car may now be passing 
through an unusually rapid phase in its recovery from the 19th-century 
Great Eruption.  Independent of this conjecture, our data show
that the 5.5-year cycle is not a straightforward clocklike affair;   
while the emission-line behavior reported here is essential for 
modeling the spectroscopic event.  Parallel ground-based observations 
of the polar-latitude spectrum reflected by the ``Homunculus'' 
ejecta-nebula will be described in a companion paper by \citet{ws05}.

\section{ THE OBSERVATIONS }  

In this paper the name $\eta$ Car denotes only the central star 
and its wind, excluding ejecta at radii larger than 0.1{\arcsec} 
or about 200 AU.   During 1998--2003 we observed it with the 
HST/STIS/CCD on 21 occasions, about half of them during 2003. 
Table 1 lists the observation dates specifically mentioned in this 
paper;  for more information see {\tt http://etacar.umn.edu},
and/or the HST archives at {\tt http://stsci.edu}.
The spectroscopic cycle has period =   
2023 $\pm$ 7 d $\approx$ 5.54~y, based on photometry, spectroscopic 
variations, and X-ray observations \citep{sa01,pw04,ad00,corc04}.   
Since the circumstellar extinction has diminished rapidly in recent
years, causing a dramatic increase in the star's apparent brightness 
\citep{kd99b,mk04}, we measure emission line profiles relative 
to the underlying continuum.  Lines and continuum all 
originate in the opaque stellar wind, not at the star's surface 
\citep{kd87,kd95,dh97,jdh01}.


Each spectrum discussed below represents essentially a 0.1{\arcsec}
$\times$ 0.15{\arcsec} spatial sample:   the pixel size was about
0.05{\arcsec}, the slit width was about 2 CCD columns, each
extraction sampled about 3 CCD rows, and the spectral resolution
was roughly 40 km s$^{-1}$.  (Here we omit some technicalities  
that have little effect on the main results.)   For the 
bright H$\alpha$ and H$\beta$ lines we use only data with 
integration times less than 1 s and 20 s respectively, short enough 
to avoid saturation of any CCD pixels; and we have verified that 
in fact none were saturated.  Other instrument parameters and our 
non-standard reduction techniques, inessential for this paper and 
too complex to describe briefly,  can be found at 
{\tt http://etacar.umn.edu}
and will be reviewed in a series of later, more detailed papers.
Doppler velocities quoted below are all heliocentric.

\section{ BEHAVIOR OF THE HYDROGEN EMISSION LINES } 


  Figure 1 shows how H$\alpha$, the brightest observable feature, 
varied during 1998--2003.  The conspicuous absorption near 
$-150$ km s$^{-1}$ must occur far outside the stellar wind, since
it appears in STIS data across much of the inner Homunculus Nebula.   
Its central wavelength varies with position, in a way that differs
substantially from the net Doppler shift of the star's spectrum
reflected by dust in the expanding ejecta.  This absorption 
feature fluctuates in strength and confused earlier ground-based 
observations;  see Fig.\ 4 in \citet{kd99a}, Fig.\ 2 in \citet{mcg99},
or Fig.\ 4 in \citet{rmh02}.   Unfortunately it masks part of 
the intrinsic line profile.  

  Apart from the extraneous narrow absorption, during the 1997--98 
spectroscopic event H$\alpha$ resembled a conventional stellar-wind
emission feature with a rounded peak, P Cyg absorption around 
$-500$ km s$^{-1}$, and extended wings probably due to Thomson 
scattering (Fig.\ 1a).  \citet{jdh01} described a wind model that 
accounted for the early-1998 spectrum fairly well; see also 
\citet{kd99a}.    By 1999, however, P Cyg absorption had almost 
disappeared along our line of sight to the star (Fig.\ 1b).  
Meanwhile the P Cyg feature remained prominent at polar latitudes, 
observed via reflection in the southeast Homunculus lobe \citep{ns03}.  
Most likely this difference indicates that the stellar wind between 
events is denser near the poles than it is at lower latitudes.  If so,
then $\eta$ Car's photosphere, located in its wind, usually has a prolate 
shape (see also \citet{vb03}) -- except, perhaps, during a spectroscopic 
event.  The line profile varied with no clear trend or pattern during 
1999--2002 (Fig.\ 1b).  Naturally we expected H$\alpha$ to 
resemble its early-1998 form during the 2003.5 event.

   Fig.\ 1c shows that in fact it behaved quite differently.  
Our 2003.51 and 2003.58 observations bracketed the time when
5.54 years had elapsed since the earliest STIS data; 
in order to simplify the figure we omit the 2003.51 profile,
which closely resembled 2003.58 but was slightly higher.
Comparing the 2003.58 data to 1998.00, one finds that --- 
 \\
 $\bullet$  The line center at 2003.58 had a remarkably flat top 
 with $F_{\lambda} \approx 42 F_{\lambda}({\rm continuum})$, 
 only about half the relative height seen in 1998. 
 \\ 
 $\bullet$  The long-wavelength wing in 2003 was about 30\% lower 
 than in 1998.  (Due to the Thomson scattering process, this is
 expected to be relatively insensitive to the viewing direction.)  
 \\  
 $\bullet$  The short-wavelength side and the P Cyg feature behaved 
 more or less the same in 2003 as they had in 1998 -- except, perhaps,
 that P Cyg absorption appeared deeper at 2003.58 than it
 did at 1998.00.  
 \\
 $\bullet$  The total H$\alpha$ equivalent width at 2003.58 was 
 550 {\AA}, compared to 880 {\AA} at 1998.00 -- nearly a 40\% decrease.   
 \\ 
 $\bullet$  In summary, the flux temporarily decreased at Doppler
 velocities around zero during the 2003 spectroscopic event, but
 not in the preceding one.   
 

The 2003.37 and 2003.72 curves in Fig.\ 1c illustrate a more 
subtle trend that occurred around the time of the event.   
Several weeks before 2003.5 the maximum flux occurred at negative 
Doppler velocities;  then, between 2003.5 and 2003.7, the top of
the profile tilted smoothly so that afterward the maximum was on 
the positive-velocity side.   If one were to measure the line centroid 
during the entire 5.5-year cycle (somehow allowing for the extraneous 
absorption near $-150$ km s$^{-1}$), then a systematic fluctuation 
would appear, resembling  those shown in \citet{ad97} and \citet{ad00}; 
but this involves complex variations in the line shape, 
and does not represent a classical orbital velocity curve. 
(See also \citet{kd97,kd99,kd00b} and other refs.\ cited therein.)  
We shall return to this point in Section 5 below.

By 2003.88 the H$\alpha$ profile had recovered to a more normal 
appearance (Fig.\ 1d), strongly suggesting that the flat top 
was related to the spectroscopic event.\footnote{
   Later STIS observations in March 2004 showed a profile much 
   like 2003.88, but slightly stronger. }
We are not aware of any observation of a flat-topped H$\alpha$ profile 
in previous direct spectroscopy of $\eta$ Car, including ground-based
data obtained during the weeks preceding the 1998.0 STIS observations
(see \citet{mcg99}).   Broad flat-topped line shapes can be seen, however, 
in some ground-based observations of the southeast Homunculus lobe, 
reflecting a polar view of the wind \citep{riv01,ws05}.

A line profile of this type is not, {\it per se,\/} very surprising. 
It is well known that an idealized, optically thin expanding shell 
produces rectangular line shapes, and the 
``optically thin'' proviso is not very strict;  a flat-topped profile 
occurs whenever the observed emission represents the vicinity 
of a quasi-spherical expanding surface.  This situation can arise in 
an ejected shell, or, alternatively, near a localized abrupt gradient 
in ionization, excitation, or (as an anonymous reviewer suggests) 
continuum opacity.  Flat-topped line profiles can also result from 
very different configurations, e.g. from a pair of oppositely-directed 
polar flows viewed at an oblique angle.  In this paper we do not 
advocate any one model in particular.  Rather, we wish to emphasize 
a basic puzzle and surprise:  {\it Flat-topped hydrogen line 
profiles characterized the 2003 spectroscopic event but not the 
preceding one in 1997-98.\/}


  Figure 2 shows three other hydrogen lines in 1998 and 2003.
H$\beta$ and P$\eta$ behaved qualitatively like H$\alpha$.    
In the 2003.58 data H$\beta$ had a flat top at about 
$12 F_{\lambda}({\rm continuum})$, while its extraneous 
absorption near $-150$ km s$^{-1}$ was weak because this line
has a smaller oscillator strength than H$\alpha$.   
On the other hand H$\eta$, the high-order Balmer line 
which is least confused with adjoining features, did not show 
much difference between 1998.00 and 2003.58;  on both occasions 
it resembled a compromise between the 1998 (high-peaked) and 
the 2003 (flat-topped) shapes seen in H$\alpha$ and H$\beta$.  
In wind models like that described by \citet{jdh01}, H$\eta$ 
originates at smaller radii than do H$\alpha$ and H$\beta$.
The Paschen lines are somewhat more difficult to interpret 
in this object \citep{kd00b}.  Emission lines of Fe~II and He~I 
varied in more complex ways, beyond the scope of this paper.


  On three occasions before the STIS became available, the HST's Faint 
Object Spectrograph (FOS) was used to obtain spectra of $\eta$ Car with 
spatial resolution better than 0.3{\arcsec} \citep{kd95,rmh99a}.   
Since H$\alpha$ was too bright to observe with the FOS,  Table 2
lists the equivalent width of H$\beta$ from 1991 to 2003.  By a 
significant margin, the smallest observed values occurred during 
the 2003 event.   


Figure 3 shows how the total H$\alpha$ emission and its P Cyg absorption
varied with time.  The 2003 event provided the following clues regarding
H$\alpha$ and H$\beta$:   \\
$\bullet$  Some developments occurred rapidly.  For instance, 
as P Cyg absorption deepened, the H$\alpha$ flux at Doppler velocities 
near $-525$ km s$^{-1}$ declined from $5.1 F_{\lambda}(\rm continuum)$ 
to $2.8 F_{\lambda}({\rm continuum})$ in 21 days, and then fell to 
only $0.7 F_{\lambda}({\rm continuum})$ in the next 12 days.  \\   
$\bullet$  Other characteristics appeared earlier 
and developed more gradually.   The decrease in
total H$\alpha$ emission was well underway at 2003.24, 
three months before the rapid  stage of the event.  
(Note the smooth turnovers soon after 2003.0 
in both parts of Fig.\ 3.)  \\ 
$\bullet$  The P Cyg absorption feature did not vanish as abruptly 
as it had deepened;  it remained substantial five months after 
the rapid phase mentioned above.    (In Fig.\ 1d, compare 
2003.88 to 2003.12 at velocities around $-500$ km s$^{-1}$).  \\    
$\bullet$  The FOS and STIS data suggest, albeit uncertainly,  
that H$\alpha$ and H$\beta$ may tend to be brightest between
1.5 and 0.5 years before an event.

Some of these remarks have parallels in other reported observations, 
such as near-infrared photometry \citep{pw04}, helium line strengths
seen in ground-based data \citep{ad99}, and the X-ray flux 
\citep{corc04}.   Spectroscopy of the type reported here is especially 
valuable, however, because it is more specific;  it samples larger
numbers of observables and structural parameters of the wind than 
X-rays or photometry do.   (Photometry, moreover, is affected by
varying circumstellar extinction, while the observable 2--10 keV 
X-rays probably represent only the hottest parts of a localized 
shock structure, almost independent of the large-scale wind.)  

In summary, we have found an element of paradox in the 1998--2003 
spectroscopic cycle:   the H$\alpha$ profile was {\it least\/} 
flat-topped during the 1997--98 event, but then was {\it most\/} 
flat-topped during the 2003 event (Fig.\ 1).  Since it partially 
recovered after 2003.7,  this is not simply a long-term trend.

\section{ RECENT ABNORMALITIES }  

Whenever moderate changes are seen in $\eta$ Car, one is tempted to 
ascribe them to routine LBV-style outbursts and fluctuations like those 
reviewed by \citet{hd94}.  In this case, however, two circumstances 
make that view unappealing. 

  First, the minimum intensity and maximum flattening of H$\alpha$ 
coincided perfectly with the mid-2003 spectroscopic event 
(Figs.\ 1c and 3).  If the previous event in 1997--98 had not been
observed, we would now assume that this behavior is merely a detail 
of the 5.5-year cycle.  In fact, however, the earlier STIS data and
ground-based spectroscopy reported by \citet{mcg99} show that 
no similar development occurred in 1997--98.
Other emission lines also differed between 1998.0 and 2003.5, 
but they are more complex and will be discussed in a later, more 
detailed paper.  Admittedly each spectroscopic event may be sensitive 
to irregular underlying LBV-like fluctuations, but a second anomaly 
also occurred during the same period:  $\eta$ Car has exhibited 
unusual photometric behavior since 1997.

  In 1998 and again in 2003, the star's apparent brightness increased 
at an extraordinary rate and did not appreciably subside later 
\citep{kd99b,mk04,pw04}.  Two facts indicate that this was not 
merely a normal aspect of the 5.5-year cycle.  First, it coincided 
with the most extreme rise in the ground-based photometric record 
of the past 50 years (see remarks in next paragraph).
Second, the central star has brightened by a factor too large 
to have occurred repeatedly before 1998;  if it continues at 
the rate observed with STIS,  then after only two or three 
more 5.5-year periods the star will appear nearly as bright 
as it did 200 years ago, when there was little circumstellar 
extinction.\footnote{  
   This remark should not be construed as a prediction,
   though such a development is quite possible.  Our point
   is that the recent brightening was, indeed, a
   major development.  }

Some authors -- \citet{st99,vg03,pw04} -- interpret the recent 
brightening as a relatively ordinary phenenon, perhaps a normal 
``LBV S Dor phase,'' but such a view is difficult to sustain even 
if one ignores the HST photometry.  It is true that ground-based 
photometry showed one or two earlier examples of rapid brightening, 
especially around 1981 when the Homunculus brightened by 0.3 magnitude 
at visual wavelengths.  In some published figures, the 1981  
episode appears more conspicuous than the 1998--99 increase 
merely because it had more data points; see Fig.\ 2b in 
\citet{kd99b} and Fig.\ 7 in \citet{mk04}.  However, soon 
after 1981 the brightness subsided by nearly the same 
amount, so there was no departure from the longer-term 
trend line.  The 1996--2004 visual-wavelength record shows
a larger increase, while -- more important --
the brightness has not returned to ``normal.''  In 2003 it was 
still almost 0.3 magnitude above the average 1952--1992 trend 
line.  This has not been an ordinary LBV or S Dor eruption, 
since the spectrum has not changed appropriately.  Moreover, the 
above statements refer mainly to the Homunculus, while 
our HST photometry of the central star shows a far more 
dramatic increase since 1997 (see below).  This must be a
recent development, not a continuation of a long-term trend,
since an extrapolation of the 1997--2003 rate of change back 
to about 1985 would ``predict'' that the central star should 
have been fainter than the Weigelt condensations at that time, 
contrary to speckle observations reported by \citet{we86} and 
\citet{hw88}.  In other words, the rate must have accelerated 
sometime in the 1990's.  Altogether, then, $\eta$ Car's brightening 
since 1997 has {\it not\/} been like anything else in the modern 
photometric record.   Something similar may have occurred in 
1938--1952 \citep{dev52,djk56}, but not more recently.

Some interpretive details are essential.    During 1998--2003 
the apparent red-wavelength brightness of the central star 
tripled according to HST/STIS observations \citep{mk04}, and 
the visual-wavelength increase was probably larger.  Since the
star is close to the Eddington limit and has shown only 
moderate spectral changes, this development cannot have been 
entirely an intrinsic brightening;  it must have included 
a rapid decrease of the circumstellar extinction.  For reasons 
noted by \citet{kd99b}, this almost surely indicates a real 
decrease in the amount of dust near the star ($r < 2000$ AU), 
and not, e.g., movement of a dusty condensation that intercepts 
our line of sight.  Either existing dust was destroyed, or the 
dust formation rate decreased in the mass outflow a few hundred 
AU from the star, or both.   Plausible causes involve changes 
in the stellar wind and/or its emergent radiation field. Therefore 
{\it it is not easy to explain the recent brightening without 
invoking some change in the star itself.\/}   

\section{ IMPLICATIONS }  

  Thus, two concurrent developments -- a striking difference between 
the 1998.0 and 2003.5 spectroscopic events, and the unprecedented 
brightening in 1998--2003 -- lead us to suspect that {\it $\eta$ Car's 
outer layers have recently changed at an unusually rapid rate.\/}  
This possibility is consistent with certain earlier peculiarities 
in the historical record, and with the physical state of the object.

This star's thermal and rotational structure has been gradually 
recovering from the Great Eruption that occurred in the 1840's 
\citep{dh97,rmh99b,kd00a, rmh02,ns03}.  One might assume that today,
160 years after the eruption, the star has approached a quasi-steady 
state, and that the long-term photometric brightening is caused 
only by expansion of dusty ejecta.  But such a guess appears to be 
incorrect:  a pronounced systematic He~I $\lambda$10830 decrease 
observed since 1980 \citep{ad99}, the spectroscopy reported in this 
paper, and other factors all indicate that the structure of the 
wind continues to evolve.  Therefore, presumably, so does the outer 
structure of the star, and the overall timescale seems theoretically
reasonable \citep{mae05}.  Lacking exact models for the readjustment 
process, we should not be very surprised if the luminosity, radius, 
and wind parameters occasionally exhibit abnormal rates of change.  
\citet{rmh99b} and \citet{rmh05a} have noted that four remarkable 
episodes -- the Great Eruption, the second eruption in the 1890's, 
a mysterious photometric-spectroscopic change during the 1940's 
(see below), and the recent brightening -- were separated by roughly 
50-year intervals.  Even if this timescale proves illusory, the 
developments just named were clearly real.

For example, evidence indicates that the present-day form of the 
5.5-year cycle first appeared sometime in the 1940's.  \citet{sa01} 
have found that spectrograms from the years 1899--1919 show no sign 
of He~I emission at the level now usually observed in $\eta$ Car, and 
\citet{rmh05b} has extended the same statement to 1941 based on
spectrograms in the Harvard plate collection.  Apparently, during 
that entire period the stellar wind was in its ``low-excitation'' 
state, now seen only during a spectroscopic event.  A high-excitation 
state appeared, however, sometime before the early 1950's \citep{gav53}.  
Meanwhile the photometric behavior changed dramatically during the 
same era.   Following about 40 years of near-constant apparent 
magnitude, $\eta$ Car brightened significantly between 1938 and 1941 
\citep{djk56};  then \citet{dev52} observed an extremely rapid 
jump,  after which the consistent photometric trend of 1952--1992 began.  
Therefore the photometric trend and the spectroscopic state both changed 
conspicuously in an interval of less than 15 years.  The role of that 
episode in the current binary vs.\ single-star puzzle is unclear, but 
a change in the observed stellar wind almost certainly occurred.  Most 
likely the mass-loss rate was substantially larger before 1938, or 
at least a lower-density zone at low latitudes \citep{ns03} did not 
exist then.  The 1938--52 change may have been the stage of recovery 
when the surface rotation rate had spun up to a rate sufficient for 
the polar and equatorial wind zones to separate.  Regardless of the 
precise explanation, the record of those years suggests that a new 
and roughly analogous development might occur again.

Certain ground-based observations support the idea that a rapid 
development is now underway.  During the past 20 years the He~I 
$\lambda$10830 emission has shown a clear secular trend toward 
lower intensity \citep{ad99}; indeed those authors' Figure 2 
seems to foretell a culmination of the trend in the near future.  
Some parameters of the 5.5-year cycle in near-IR photometry 
\citep{pw94} may also show progressive tendencies, admittedly 
less definite than the He~I $\lambda$10830 behavior.

In summary:  The STIS spectroscopy reported in this paper, HST and 
other photometry, Damineli's He~I $\lambda$10830 observations, the 
1938--52 precedent, and additional factors together inspire our 
suspicion that $\eta$ Car is now in an usually rapid stage of its 
post-eruption thermal and rotational recovery.  We do not claim 
to have proven this conjecture, and quantitative models are 
obviously needed;  but the qualitative hypothesis appears 
coherent, plausible, and interesting.

Without theoretical models far beyond the scope of this paper, one 
cannot make a detailed forecast.   The star may become more 
stable during the next 10 or 20 years, as it probably was before the 
year 1700, perhaps with a decreased mass-loss rate.  Less likely, the 
wind might instead increase, as it did between 1750 and 1830 
\citep{dh97,rmh02}.  In either case we should not be surprised if 
the latitude structure discussed by \citet{ns03} evolves on the same 
time scale, related to the surface rotation rate.  Assertions like these 
are not qualitatively new, but we suggest that the relevant time scale 
may currently be only 10 or 20 years rather than more than 40 years as 
usually assumed in the past.   Even if the rate of brightening abates 
somewhat from its recent average, by about the year 2020 the central 
star will have become brighter than the Homunculus Nebula and their 
total will be of 4th apparent visual magnitude -- in other words, 
to the unaided eye $\eta$ Car will appear much as it did to Halley 
more than three centuries ago.

Independent of the conjecture expressed above, the HST/STIS data
reported in Section 3 provide valuable information on the nature 
of a spectroscopic event, unmatched by any existing ground-based data.    
The H$\alpha$ profile seen in mid-2003 is difficult to interpret 
because it represents a complex morphology, not necessarily spherical.  
The main difference between the 1998.0 and 2003.5 profiles (Fig.\ 1) 
is that material with Doppler velocities near zero, either 
slow-moving gas or else mass flow near the plane 
of the sky, became less conspicuous during the 2003 event.  The 
absence of a central line peak accounts for most of the decrease 
in total emission.

Several rival explanations of $\eta$ Car's spectroscopic events
are available:
\\ $\bullet$ Each event may be primarily a mass-ejection episode
{\it a l\`{a}\/} \citet{zanella84}, or at least a global disturbance 
in the wind geometry.  The ejection morphology may be shell-like or 
toroidal, and the event may be triggered by the close approach of 
a hypothetical companion star -- although the latter detail is 
not a necessary part of this scenario. 
\\ $\bullet$  Alternatively, temporary spectroscopic changes of the 
same type may result if a hot secondary star moves behind the primary
wind, so that its UV photons cannot excite the side of the
wind that faces us.  (This is not, properly speaking, an
eclipse, since the hypothetical secondary star does not directly
contribute much to the brightness at any time.)  
\\ $\bullet$  Or, perhaps, the excitation and ionization of visible
parts of the primary wind may decrease when a hot secondary
star's ultraviolet photons become engulfed, near periastron, 
in the primary's dense inner wind.  
\\  Other possibilities may exist but these are the most obvious. 
They are merely {\it ideas,\/} not models;  none of
them is straightforward, none of them has been calculated, and they 
are not mutually exclusive \citep{kd99}.   Unfortunately the observations
reported here do not provide an immediate, simple qualitative test,
As we noted in Section 3, flat-topped emission lines seem especially
natural in an expanding shell;  but this remark does not 
prove the other two ideas wrong.  Instead of advocating a particular
scenario, here we draw attention to two points:  (1) The line profiles
and flux variations observed with STIS provide essential constraints
for realistic models, and (2) {\it one must ask why the 1997-98 event
did not exhibit the same emission-line behavior.\/}

Several other aspects of our data merit consideration.  For instance, 
the line peak shifted from negative to positive velocities during 
the 2003 event -- e.g., compare 2003.37 to 2003.72 in
Fig.\ 1c.  In a binary-star model this may be a clue to
the orbit orientation.  In almost every proposed binary
scenario, the 5.5-year orbit is highly eccentric ($e > 0.6$) 
and a spectroscopic event occurs near periastron.    
Suppose the  major axis is roughly perpendicular to our 
line of sight, and in mid-2003 the hot companion star was near 
periastron and moving away from us.  In that case,  before and 
after periastron it would have tended to excite material, 
respectively, on our side of the primary wind and then on 
the other side -- emphasizing negative and then positive velocities 
as observed.  The same arrangement can account for the fact that 
H$\alpha$ P Cyg absorption, the X-ray flux, and some other observables 
change rapidly just before the event reaches its maximum, and 
recover more gradually afterward.  This orbit orientation has 
been advocated by \citet{bish01} and \citet{kd02a} regarding 
the X-ray behavior, but it is quite different from that proposed 
by, e.g.,  \citet{ad00} and \citet{pc02}.  {\it Caveat:\/}
Like most discussions of binary models for $\eta$ Car, these
remarks are highly speculative and one can easily 
invent alternatives \citep{kd99}.


The hydrogen lines help to define accurately
when the 2003 event occurred.  Figure 4 shows the
behavior of six observables representing $\eta$ Car's
wind;  evidently the best time-markers for the event
were the near-infrared flux\footnote{
   If we use ground-based photometry for this purpose, 
   then near-IR wavelengths seem better than visual 
   wavelengths, being somewhat less dominated by light 
   reflected in the Homunculus. }
(Fig.\ 4b), the H$\alpha$ P Cyg absorption (Fig.\ 4d), and perhaps 
the He~I $\lambda$6680 flux (Fig.\ 4e).  
The X-ray flux also changed rapidly (Fig.\ 4f), but in 
an erratic manner that did not match the curve seen in 
late 1997 \citep{corc04}.   The final deepening of the 
P Cyg feature occurred close to MJD = 52820, several
days before the midpoint of the H-band photometric
drop reported by \citet{pw04} -- i.e., at the 
beginning of July 2003 with an uncertainty
of only a few days.  Unfortunately, it is not  
clear that comparable observations will be attempted
during the next event in early 2009.  The HST/STIS
will not be available then, while, 
ironically, ground-based work is also becoming more 
difficult due to the shortage of appropriate small 
telescopes.  On the other hand, if the central star 
continues to grow in brightness relative to the
Homunculus Nebula, adequate visual-wavelength 
spectroscopy of it will become feasible with 
ground-based equipment.

Figures 3 and 4 suggest two characteristic time-scales for an 
event:  100--200 d for some parameters, and 5--20 d for the 
most rapid changes.   The former might be the 
quadrature-to-periastron time for an eccentric  binary orbit,
or the star's rotation period, or the thermal/dynamical
timescale for growth of an instability in $\eta$ Car's
outer layers.  The latter, shorter timescale seems a 
``natural'' value for a disturbance in the wind:  
500 km s$^{-1}$ $\times$ 10 d $\approx$ 3 AU, 
which is the order-of-magnitude characteristic size
for the inner wind and also for the periastron of an 
eccentric binary orbit with a 5.5-year period.  The 
star's dynamical timescale is also of the same order
of magnitude.

We emphasize again that our main point is practically 
model-independent:   The 1997-98 and 2003 events differed in 
respects which probably indicate a serious change in the 
stellar wind structure.  The detailed line profiles will 
be valuable for constraining models, but that, in a sense, 
will be a second-order application of the data.  
The complex behavior of He~I, Fe~II, and other spectral
features of the stellar wind observed with HST/STIS,
as well as spectroscopy of ejecta at radii $>$ 0.1{\arcsec}, 
will be reported in a series of future 
papers.  Finally, we caution that all observations and 
most comments in this paper refer only to our direct
view of the star.  Reflection in the Homunculus
Nebula allows an indirect ``polar'' view of $\eta$ 
Car, with potentially different emission line profiles.  
A preliminary report on that view, employing the
ground-based VLT/UVES instrument, will be given 
in a companion paper by \cite{ws05}.

--- --- ---

{\it Acknowledgements\/}

We are grateful to Beth Perriello at STScI for extensive assistance 
in planning the Treasury Project observations, and to Matt Gray and
Michael Koppelman at the University of Minnesota for helping with 
non-routine steps in the data preparation and analysis.  We thank
an anonymous reviewer for drawing our attention to several points 
where the original version of this paper was insufficiently clear.  

The HST Treasury Project for Eta Carinae is supported by NASA 
funding from STScI (programs GO-9420 and GO-9973), and in this paper 
we have employed HST data from several earlier GO and GTO programs.  
Treasury Project CoI's who did not participate in this particular paper 
are S.\ Johansson, N.\ Walborn, M.\ Bautista, and H.\ Hartman.  

\clearpage




\clearpage 

\begin{figure}
\epsscale{0.5}
\plotone{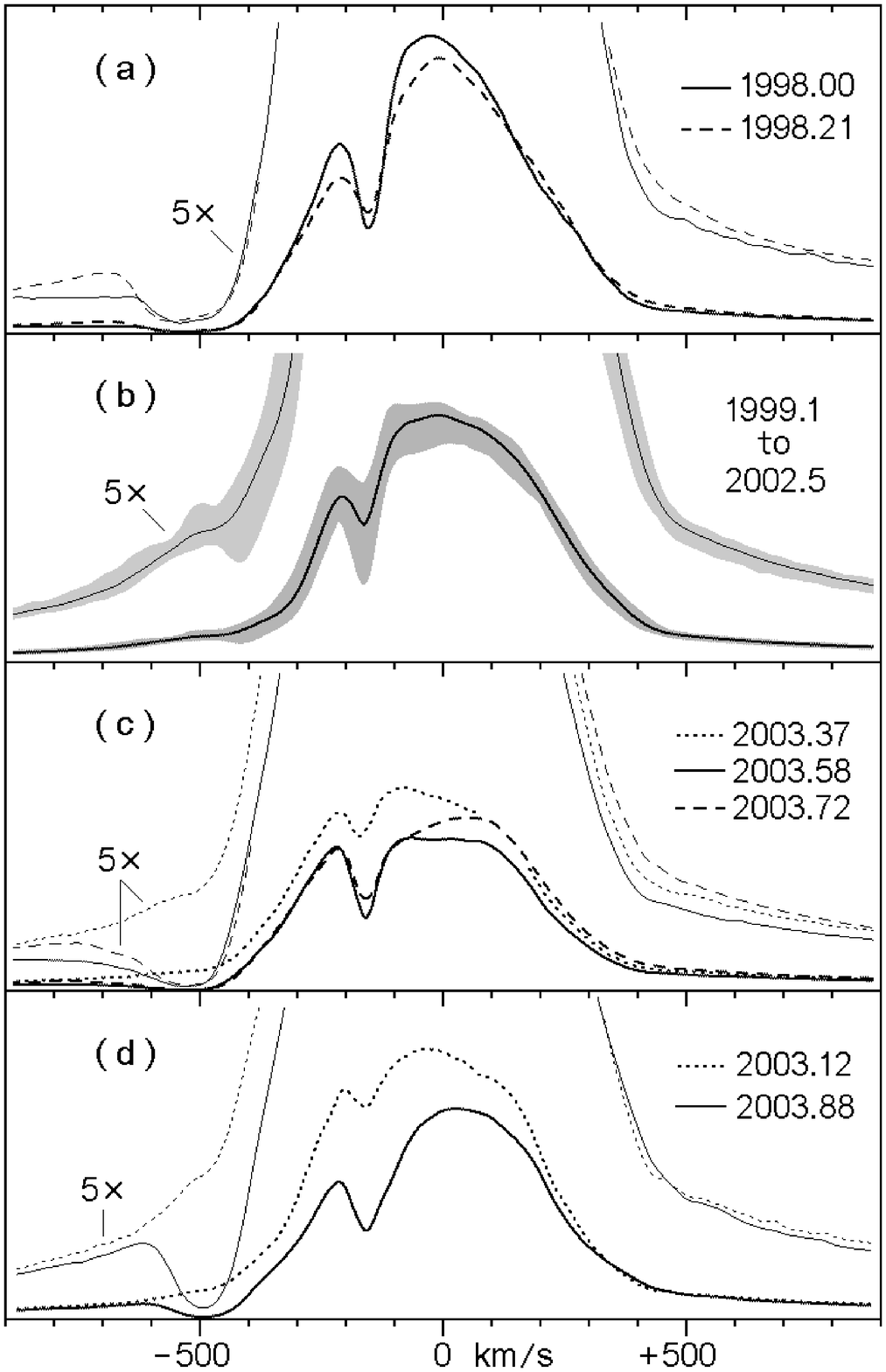}       
\caption{ Profile of $\eta$ Car's stellar-wind H$\alpha$ emission line 
   during 1998--2003.  The bottom of each subfigure is at zero flux,
   while the top represents $90 F_{\lambda}({\rm continuum})$ for 
   the main curves or $18 F_{\lambda}({\rm continuum})$ 
   for lighter curves marked ``$5 {\times}$''; 
   we refer to the average continuum in the wavelength range 
   6740--6800 {\AA}.  Subfigure 1b shows the average and the envelope 
   of six observations in 1999--2002, the less active portion of the 
   5.5-year spectroscopic cycle.  As noted in the text, the
   prominent absorption near $-150$ km s$^{-1}$ occurs outside 
   the stellar wind.  These curves have not been smoothed;
   the statistical noise is relatively small. }
\end{figure}

\clearpage

\begin{figure} 
\epsscale{0.5}
\plotone{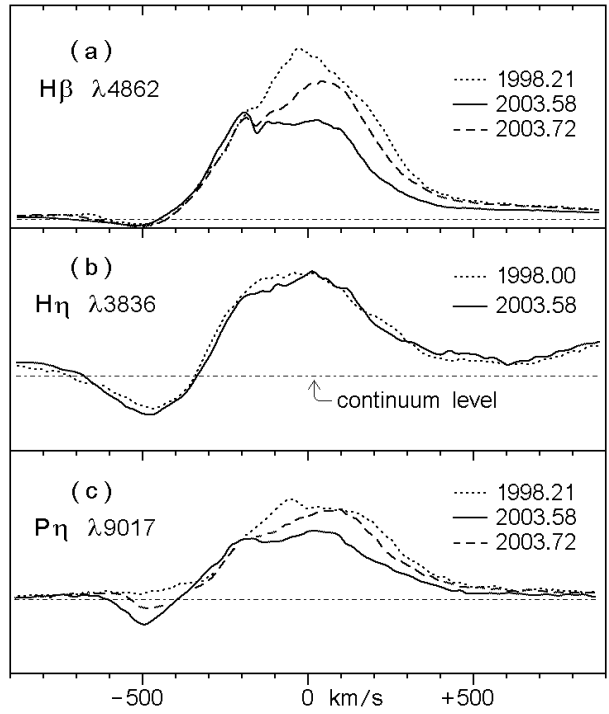}
\caption{ Profiles of the Balmer $\beta$ (2--4), Balmer $\eta$ (2--9),
   and Paschen $\eta$ (3--10) emision lines near the 1998 and 2003.5 
   events.  H$\beta$ and the Paschen lines were not observed with STIS 
   at 1998.00, because insufficient HST time was available on that
   occasion.  The top boundary represents a flux of 
   $25 F_{\lambda}({\rm continuum})$ in subfigure 2a and 
   $3 F_{\lambda}({\rm continuum})$ in 2b and 2c. }   
\end{figure}

\clearpage

\begin{figure}
\epsscale{0.6}
\plotone{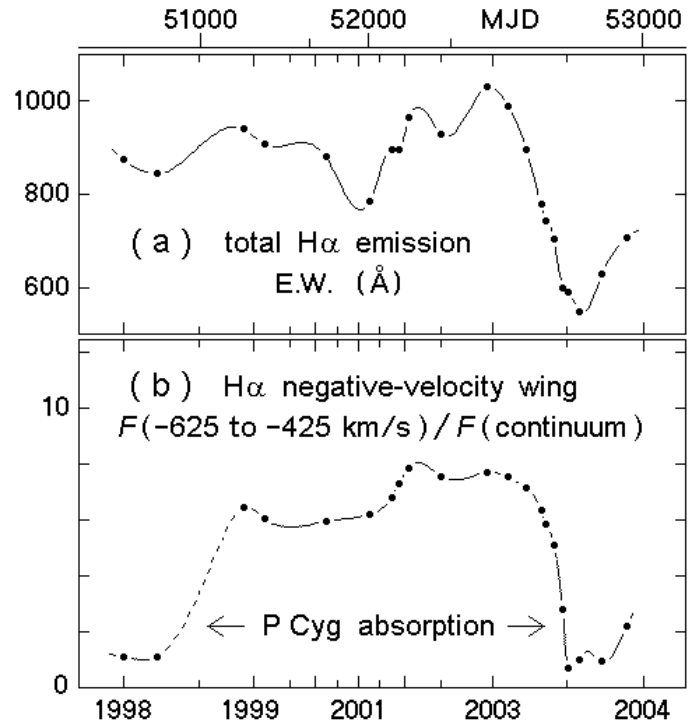}
\caption{(a) Equivalent width of H${\alpha}$ emission and (b) average 
   flux in the wavelength interval where H${\alpha}$ P Cyg absorption
   occurs, as functions of time.  The time scale is systematically 
   distorted in order to show the 1998 and 2003 events more clearly.  
   Data points are connected by spline interpolations as an aid to 
   viewing the progression;  the important mid-1998 interval is, 
   therefore, merely conjectural.}
\end{figure}  

\clearpage

\begin{figure}
\epsscale{0.5}
\plotone{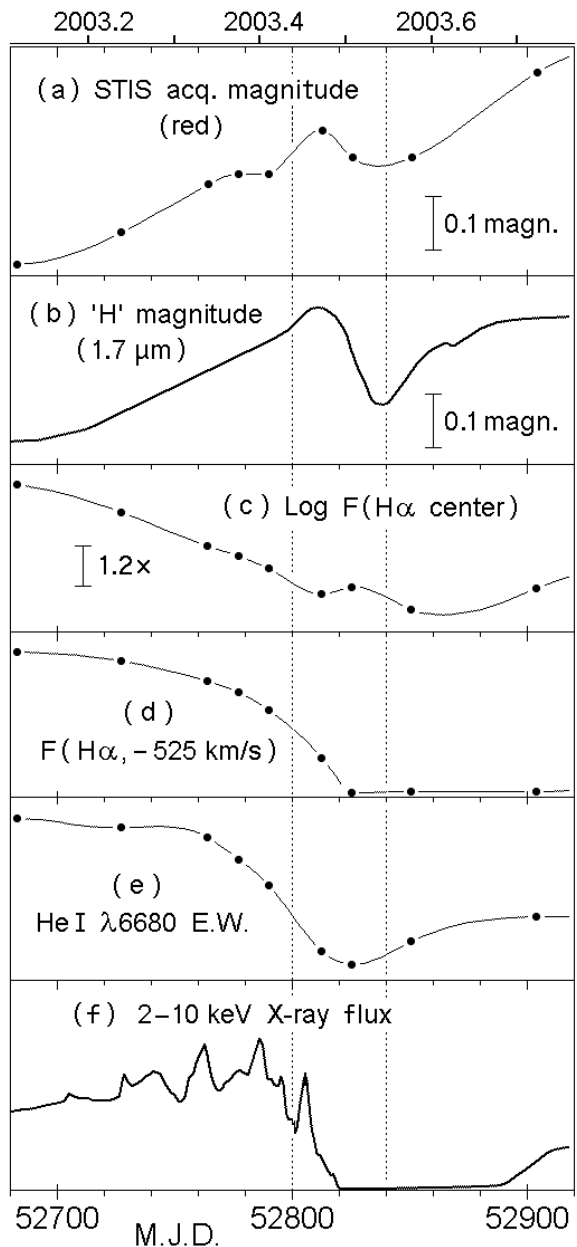}
\caption{ Observables of the stellar wind that can be used as
  time-markers for the 2003.5 spectroscopic event.  
  (a) The red-wavelength brightness observed with the HST/STIS
  \citep{mk04};  (b) the near-IR, ``free-free'' continuum
  brightness \citep{pw04}; (c) flux at the H$\alpha$ peak;  
  (d) flux at the average wavelength of P Cyg absorption;
  (e) equivalent width of the He~I $\lambda$6680 emission
  line;  (f) X-ray flux observed with the RXTE satellite
  \citep{corc04}.  Subfigures (a) and 
  (c)--(e) are based on our HST/STIS observations.
  (a)--(c) are logarithmic plots while (d)--(f) are linear
  with zero level at the bottom of each subfigure.  Two 
  light vertical lines mark convenient reference
  times, MJD 52800 and 52840. }
\end{figure}  

\clearpage

\figcaption[davidson2.fig1.ps]{ 
 Profile of $\eta$ Car's stellar-wind H$\alpha$ emission line  during 
 1998--2003.  The bottom of each subfigure is at zero flux,
 while the top represents $90 F_{\lambda}({\rm continuum})$ for
 the main curves or $18 F_{\lambda}({\rm continuum})$
 for lighter curves marked ``$5 {\times}$'';
 we refer to the average continuum in the wavelength range
 6740--6800 {\AA}.  Subfigure 1b shows the average and the envelope
 of six observations in 1999--2002, the less active portion of the
 5.5-year spectroscopic cycle.  As noted in the text, the
 prominent absorption near $-150$ km s$^{-1}$ occurs far outside
 the stellar wind.  These curves have not been smoothed;
 the statistical noise level is relatively small. }

\figcaption[davidson2.fig2.ps]{ 
 Profiles of the Balmer $\beta$ (2--4), Balmer $\eta$ (2--9), and Paschen 
 $\eta$ (3--10) emision lines near the 1998 and 2003.5 events.  
 H$\beta$ and the Paschen lines were not observed with STIS
 at 1998.00, because insufficient HST time was available on that
 occasion.  The top boundary represents a flux of
 $25 F_{\lambda}({\rm continuum})$ in subfigure 2a and
 $3 F_{\lambda}({\rm continuum})$ in 2b and 2c. }

\figcaption[davidson2.fig3.ps]{
 (a) Equivalent width of H${\alpha}$ emission and (b) average flux in 
 the wavelength interval where H${\alpha}$ P Cyg absorption
 occurs, as functions of time.  The time scale is systematically
 distorted in order to show the 1998 and 2003 events more clearly.
 Data points are connected by spline interpolations as an aid to
 viewing the progression;  the important mid-1998 interval is,
 therefore, merely conjectural.}

\figcaption[davidson2.fig4.ps]{
 Observables of the stellar wind that can be used as
 time-markers for the 2003.5 spectroscopic event.
 (a) The red-wavelength brightness observed with the HST/STIS
 \citep{mk04};  (b) the near-IR, ``free-free'' continuum
 brightness \citep{pw04}; (c) flux at the H$\alpha$ peak;
 (d) flux at the average wavelength of P Cyg absorption;
 (e) equivalent width of the He~I $\lambda$6680 emission
 line;  (f) X-ray flux observed with the RXTE satellite
 \citep{corc04}.  Subfigures (a) and
 (c)--(e) are based on our HST/STIS observations.
 (a)--(c) are logarithmic plots while (d)--(f) are linear
 with zero level at the bottom of each subfigure.  Two
 light vertical lines mark convenient reference
 times, MJD 52800 and 52840. }


\clearpage

\begin{deluxetable}{ccccc} 
\tablewidth{0pt}
\tablecaption{ The Main Observation Dates Mentioned in This 
    Paper\tablenotemark{a} }  
\tablehead{ \colhead{Calendar date} & \colhead{Year number} 
    & \colhead{MJD\tablenotemark{b}} & \colhead{Phase\tablenotemark{c}} 
    & \colhead{Slit P.A.}   }
\startdata
  1998/01/01  &  1998.00  &  50814  &  0.007  & 260{\arcdeg}\\   
  1998/03/19  &  1998.21  &  50891  &  0.045  & 328{\arcdeg}\\ 
  2003/02/12  &  2003.12  &  52683  &  0.931  & 303{\arcdeg}\\  
  2003/05/17  &  2003.37  &  52776  &  0.977  &  38{\arcdeg}\\ 
  2003/07/05  &  2003.51  &  52825  &  1.001  &  69{\arcdeg}\\  
  2003/08/01  &  2003.58  &  52852  &  1.014  & 105{\arcdeg}\\  
  2003/09/22  &  2003.72  &  52904  &  1.040  & 153{\arcdeg}\\ 
  2003/11/17  &  2003.88  &  52961  &  1.068  & 218{\arcdeg}\\  
\enddata
\tablenotetext{a}{STIS data were also obtained on more than
   10 other occasions during 1999--2003, not individually
   used in this paper.}
\tablenotetext{b}{Modified Julian Day = (Julian Day Number) $-$ 2400000.5.} 
\tablenotetext{c}{ Relative phase in the spectroscopic cycle, assuming
   $P$ = 2023 d = 5.54 y and arbitrarily adopting zeropoint $t_{0}$ = 
   MJD 50800 $\approx$ 1997.96.   Some other authors have used nearly
   the same $t_{0}$, which, however, does not represent any known physical 
   reference point in the cycle. } 
\end{deluxetable}

\clearpage

\begin{deluxetable}{ccc} 
\tablewidth{0pt}
\tablecaption{ Equivalent Width of H${\beta}$ in ${\eta}$ Car's Wind 
   Spectrum  Observed with the HST, 1991--2003 }
\tablehead{ \colhead{Date} & \colhead{E.W.\ ({\AA})\tablenotemark{b} } 
    & \colhead{Instrument} }  
\startdata
  1991.62   &   194   &   FOS   \\
  1996.51   &   239   &   FOS   \\
  1997.10   &   197   &   FOS   \\  
  1998.21   &   188   &   STIS  \\
  2002.05   &   211   &   STIS  \\ 
  2002.51   &   202   &   STIS  \\ 
  2003.37   &   163   &   STIS  \\ 
  2003.51   &   122   &   STIS  \\     
  2003.58   &   114   &   STIS  \\ 
  2003.72   &   151   &   STIS  \\ 
  2003.88   &   176   &   STIS  \\  
\enddata
\tablenotetext{a}{An informal r.m.s.\ uncertainty estimate for 
   each E.W.\ listed here is ${\pm}$3\%, largely due to the 
   uncertain continuum level.}
\end{deluxetable}

\end{document}